\documentclass[a4paper,UKenglish,cleveref,autoref,thm-restate]{lipics-v2021}

\usepackage{xspace}

\newcommand{\SWAQp}{SWA\xspace}
\newcommand{\SWAQ}{\ensuremath\mathsf{SWA}}
\newcommand{\size}{\ensuremath\mathsf{size}}

\newcommand{\rank}{\ensuremath\mathsf{rank}}
\newcommand{\select}{\ensuremath\mathsf{select}}
\newcommand{\predecessor}{\ensuremath\mathsf{predecessor}}
\newcommand{\weight}{\ensuremath\mathsf{weight}}
\newcommand{\enc}{\ensuremath\mathsf{enc}}
\newcommand{\cO}{\mathcal{O}}

\def\dd{\mathinner{.\,.}}

\newcommand{\defproblem}[3]{
\smallskip
\noindent\fbox{
\begin{minipage}{0.96\columnwidth}
\vspace*{.5mm}
\textsc{#1}
\vspace*{.5mm}%

\noindent
{\bf{Preprocess:}} #2
\vspace*{.5mm}%
\noindent
{\bf{Query:}} #3%
\vspace*{.5mm}%
\end{minipage}
}
\smallskip
}

\title{Size-Constrained Weighted Ancestors with Applications}

\author{Philip Bille}{Technical University of Denmark, Denmark}{phbi@dtu.dk}{https://orcid.org/0000-0002-1120-5154}{Supported by the Independent Research Fund Denmark (DFF-9131-00069B).}
\author{Yakov Nekrich}{Michigan Technological University, Michigan, US}{yakov@mtu.edu}{https://orcid.org/0000-0003-3771-5088}{Supported by the National Science Foundation under NSF grant 2203278.}
\author{Solon P. Pissis}{CWI, Amsterdam, The Netherlands \and Vrije Universiteit, Amsterdam, The Netherlands }{solon.pissis@cwi.nl}{https://orcid.org/0000-0002-1445-1932}{Supported by the PANGAIA (No 872539) and ALPACA (No 956229) projects.}

\authorrunning{P. Bille et al.} %TODO mandatory. First: Use abbreviated first/middle names. Second (only in severe cases): Use first author plus 'et al.'

\Copyright{Philip Bille, Yakov Nekrich, and Solon P. Pissis} %TODO mandatory, please use full first names. LIPIcs license is "CC-BY";  http://creativecommons.org/licenses/by/3.0/

%TODO mandatory: Please choose ACM 2012 classifications from https://dl.acm.org/ccs/ccs_flat.cfm 


\category{} %optional, e.g. invited paper

\relatedversion{} %optional

\hideLIPIcs

\nolinenumbers %uncomment to disable line numbering

\EventEditors{Hans L. Bodlaender}
\EventNoEds{1}
\EventLongTitle{19th Scandinavian Symposium and Workshops on Algorithm Theory (SWAT 2024)}
\EventShortTitle{SWAT 2024}
\EventAcronym{SWAT}
\EventYear{2024}
\EventDate{June 12--14, 2024}
\EventLocation{Helsinki, Finland}
\EventLogo{}
\SeriesVolume{294}
\ArticleNo{1}
\begin{document}

\maketitle

\begin{abstract}
The \emph{weighted ancestor} problem on a rooted node-weighted tree $T$ is a generalization of the classic \emph{predecessor} problem: construct a data structure for a set of integers that supports fast predecessor queries. Both problems are known to require $\Omega(\log\log n)$ time for queries provided $\cO(n\text{ poly} \log n)$ space is available, where $n$ is the input size. The weighted ancestor problem has attracted a lot of attention by the combinatorial pattern matching community due to its direct application to suffix trees. In this formulation of the problem, the nodes are weighted by string depth. This research has culminated in a data structure for weighted ancestors in suffix trees with $\cO(1)$ query time and an $\cO(n)$-time construction algorithm [Belazzougui et al., CPM 2021].

In this paper, we consider a different version of the weighted ancestor problem, where the nodes are weighted by any function $\weight$ that maps each node of $T$ to a positive integer, such that $\weight(u)\le \size(u)$ for any node $u$ and $\weight(u_1)\le \weight(u_2)$ if node $u_1$ is a descendant of node $u_2$, where $\size(u)$ is the number of nodes in the subtree rooted at $u$. In the \emph{size-constrained weighted ancestor}  (\SWAQp) problem, for any node $u$ of $T$ and any integer $k$, we are asked to return the lowest ancestor $w$ of $u$ with weight at least $k$. We show that for \emph{any rooted tree} with $n$ nodes, we can locate node $w$ in $\cO(1)$ time after $\cO(n)$-time preprocessing. In particular, this implies a data structure for the \SWAQp problem in suffix trees with $\cO(1)$ query time and $\cO(n)$-time preprocessing, when the nodes are weighted by $\weight$. We also show several string-processing applications of this result.
\end{abstract}

\section{Introduction}

In the classic \emph{predecessor} problem~\cite{DBLP:journals/ipl/Boas77,FW1993,DBLP:journals/ipl/Willard83,DBLP:conf/stoc/PatrascuT06,DBLP:journals/csur/NavarroR20}, we are given a set $S$ of keys from a universe $U$ with a total order. The goal is to preprocess set $S$ into a compact data structure supporting the following on-line queries: for any element $q\in U$, return the maximum $p\in S$ such that $p\leq q$; $p$ is called the \emph{predecessor} of $q$.

The \emph{weighted ancestor} problem, introduced by Farach and Muthukrishnan in~\cite{DBLP:conf/cpm/FarachM96}, is a natural generalization of the predecessor problem on rooted node-weighted trees. In particular, given a rooted tree $T$, whose nodes are weighted by positive integers and such that these weights decrease when ascending from any node to the root, the goal is to preprocess tree $T$ into a compact data structure supporting the following on-line queries: for any given node $u$ and any integer $k>0$, return the farthest ancestor of $u$ whose weight is at least $k$. Both the predecessor and the weighted ancestor problems require $\Omega(\log\log n)$ time for queries provided $\cO(n\text{ poly} \log n)$ space is available, where $n$ is the input size of the problem~\cite{DBLP:conf/esa/GawrychowskiLN14}. 

\begin{figure}[t]
    \centering
    \begin{subfigure}{0.45\columnwidth}
    \centering
    \includegraphics[width=6.5cm]{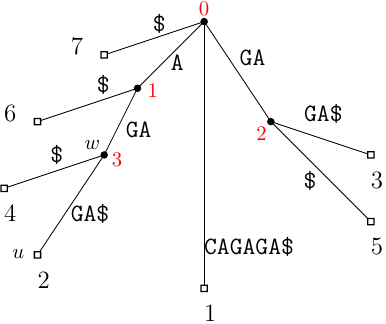}
    \caption{The internal nodes are weighted by \emph{string depth} (in red). Asking a weighted ancestor query for $i = 2$ (node $u$) and $k=2$ will take us to node $w$. Indeed, $(w,k)$ is the locus of substring $\texttt{AG}$ in the suffix tree of $X$.} 
    \label{fig:ST_WA}
    \end{subfigure}\hspace{+5mm}
    \begin{subfigure}{0.45\columnwidth}
    \centering
    \includegraphics[width=6.5cm]{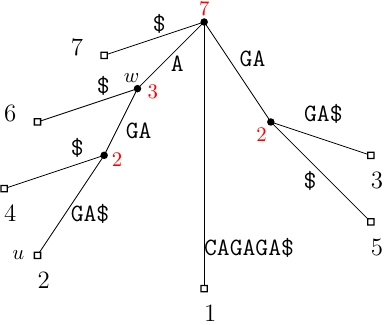}
        \caption{The internal nodes are weighted by \emph{frequency} (in red). Asking a weighted ancestor query for $i = 2$, $j = 7$ (node $u$) and $k=3$ will take us to node $w$. Indeed, $\texttt{A}$ is the longest prefix of $\texttt{AGAGA\$}$ that occurs at least $3$ times in $X$.} 
    \label{fig:ST_SCWA}
    \end{subfigure}
    \caption{Weighted ancestor queries on the suffix tree of string $X=\texttt{CAGAGA\$}$. The leaf nodes in both trees are labeled by the starting position of the suffix of $X$ they represent.}
    \label{fig:ST}
\end{figure}

The weighted ancestor problem has attracted a lot of attention in the combinatorial pattern matching community~\cite{DBLP:conf/cpm/FarachM96,DBLP:journals/talg/AmirLLS07,DBLP:conf/soda/KopelowitzL07,DBLP:journals/jda/KopelowitzKNS14,DBLP:conf/esa/GawrychowskiLN14,DBLP:conf/cpm/BelazzouguiKPR21,DBLP:conf/cpm/BadkobehCP21} due to its direct application to suffix trees~\cite{DBLP:conf/focs/Weiner73}. The \emph{suffix tree} of a string $X$ is the compacted trie of the set of suffixes of $X$; see \cref{fig:ST_WA}. In this formulation of the problem, a node $u$ is weighted by \emph{string depth}: the length of the string spelled from the root of the suffix tree to $u$; and a weighted ancestor query for two integers $i$ and $k>0$ returns the locus of substring $X[i\dd i+k-1]$ in the suffix tree of $X$. We refer the reader to~\cite{DBLP:conf/esa/GawrychowskiLN14} for several applications. This research has culminated in a data structure for weighted ancestors in suffix trees, given by Belazzougui, Kosolobov, Puglisi, and Raman~\cite{DBLP:conf/cpm/BelazzouguiKPR21}, supporting $\cO(1)$-time queries after an $\cO(n)$-time preprocessing.

However there are other tree weighting schemes that are of interest to string processing. For example, each suffix tree node can be weighted by the number of its leaf descendants; see \cref{fig:ST_SCWA}. Thus the weight of a node $u$ is equal to the frequency of the substring represented by the root-to-$u$ path.  
If we use this weighting function, then the following basic string problem can be translated into a  weighted ancestor query: 
\emph{Given a substring $I=X[i\dd j]$ of string $X$ and an integer $k>0$, find the longest prefix of $I$ that occurs at least $k$ times in $X$.}

Unfortunately, the existing data structures for the weighted ancestor problem \emph{on suffix trees}~\cite{DBLP:conf/esa/GawrychowskiLN14,DBLP:conf/cpm/BelazzouguiKPR21} depend strongly on the fact that the suffix tree nodes are weighted by string depth. They thus \emph{cannot be applied} to solve the aforementioned basic string problem.

Motivated by this fact, we introduce  a different version of the weighted ancestor problem on \emph{general rooted trees}. Let $T$ be a rooted tree on a set $V$ of $n$ nodes. By $\size(u)$, we denote the number of nodes in the subtree rooted at a node $u\in V$. Let $\weight :V\rightarrow \mathbb{N}$ denote any function that maps each node of $T$ to a positive integer, such that $\weight(u)\le \size(u)$ for any node $u\in V$ and $\weight(u_1)\le \weight(u_2)$ if node $u_1 \in V$ is a descendant of node $u_2 \in V$. The latter is also known as the \emph{max-heap property}: the weight of each node is less than or equal to the weight of its parent, with the maximum-weight element at the root. We will say that a function $\weight :V\rightarrow \mathbb{N}$ satisfying both properties is a \emph{size-constrained max-heap} weight function. For any node $u \in V$ and any integer $k>0$, a \emph{size-constrained weighted ancestor query}, denoted by $\SWAQ(u,k)=w$, asks for the lowest ancestor $w\in V$ of $u$ with weight at least $k$. The \emph{size-constrained weighted ancestor} (\SWAQp) problem, formalized next, is to preprocess $T$ into a compact data structure supporting fast $\SWAQ$ queries:

\defproblem{Size-Constrained Weighted Ancestor (\SWAQp)}{A rooted tree $T$ on a set $V$ of $n$ nodes weighted by a size-constrained max-heap function $\weight :V\rightarrow \mathbb{N}$.\\$~$}{Given a node $u\in V$ and an integer $k>0$, return the lowest ancestor $w$ of $u$ with $\weight(w)\ge k$.}

We assume throughout the standard word RAM model of computation with word size $\Theta(\log n)$; basic arithmetic and bit-wise operations on $\cO(\log n)$-bit integers take $\cO(1)$ time. 
Note that, since function $\weight$ must satisfy the max-heap property,
one can employ the existing data structures for the weighted ancestor problem \emph{on general rooted trees}~\cite{DBLP:conf/cpm/FarachM96,DBLP:journals/talg/AmirLLS07}, to answer $\SWAQ$ queries in $\cO(\log\log n)$ time after $\cO(n)$-time preprocessing (see also~\cite{DBLP:conf/spire/PissisSLL23}).
Our main result in this paper can be formalized as follows (see \cref{sec:constant_time} and \cref{sec:linear_space}).

\begin{restatable}{theorem}{theoremmain}\label{the:swaq}
For any rooted tree with $n$ nodes weighted by a size-constrained max-heap function $\weight$, there exists
an $\cO(n)$-space  data structure answering
$\SWAQ$ queries in $\cO(1)$ time. The preprocessing algorithm runs in $\cO(n)$ time and $\cO(n)$ space.
\end{restatable} 

As a preliminary step, we design an $\cO(n\log n)$-space solution using an involved combination of \emph{rank-select} data structures~\cite{DBLP:conf/wads/BaumannH19}, \emph{fusion trees}~\cite{FW1993}, and \emph{heavy-path decompositions}~\cite{DBLP:journals/jcss/SleatorT83}. We then design a novel application of \emph{ART decomposition}~\cite{AHR1998} to arrive to \cref{the:swaq}.

\subparagraph{Applications.}~Notably, Theorem~\ref{the:swaq} presents a data structure for the \SWAQp problem in suffix trees with $\cO(1)$ query time and $\cO(n)$-time preprocessing, when the nodes are weighted by a size-constrained max-heap weight function $\weight$. We show several string-processing applications of this result since $\weight(u)$ can be defined as the number of leaf nodes in the subtree rooted at $u$. Let us first provide some intuition on the applicability of Theorem~\ref{the:swaq}.

Consider a relatively long query submitted to a search-engine text database. If the database returns no (or not sufficiently many) results, one usually tries to \emph{repeatedly} truncate some prefix and/or some suffix of the original query until they obtain sufficiently many results. Our Theorem~\ref{the:swaq} can be applied to solve this problem directly in optimal time.

In particular, Theorem~\ref{the:swaq} yields \emph{optimal data structures}, with respect to preprocessing and query times, for the following basic string-processing problems (see \cref{sec:apps}):

\begin{enumerate}
    \item Preprocess a string $X$ into a linear-space data structure supporting the following on-line queries: for any $i,j,f$ return the longest prefix of $X[i\dd j]$ occurring at least $f$ times in $X$.
    \item Preprocess a dictionary $\mathcal{D}$ of documents into a linear-space data structure supporting the following on-line queries: for any string $P$ and any integer $f$, return a longest substring of $P$ occurring in at least $f$ documents of $\mathcal{D}$.
    \item Preprocess a string $X$ into a linear-space data structure supporting the following on-line queries: for any string $P$ and any integer $f$, return a longest substring of $P$ occurring at least $f$ times in $X$.
\end{enumerate}
Theorem~\ref{the:swaq} also directly improves on the data structure presented by Pissis et al.~\cite{DBLP:conf/spire/PissisSLL23} for computing the \emph{frequency-constrained substring complexity} of a given string (see \cref{sec:apps}).

\section{Preliminaries}

For any bit string $B$ of length $m$ and any $\alpha \in \{0, 1\}$, the classic \textsf{rank} and \textsf{select} queries are defined as follows:
\begin{itemize}
    \item $\textsf{rank}_{\alpha}$: for any given $i \in [1,m]$, it returns the number of ones (or zeros) in $B[1\dd i]$; more formally, $\textsf{rank}_{\alpha}(B,i) = |\{j \in [1, i] : B[j] = \alpha\}|$.
    \item $\textsf{select}_{\alpha}$: for any given rank $i$, it returns the leftmost position where the bit vector contains a one (or zero) with rank $i$; more formally, $\textsf{select}_{\alpha}(B,i) = \min\{j \in [1, m] : \textsf{rank}_{\alpha}(B,j) = i \}$.
\end{itemize}

The following result is known.

\begin{lemma}[Rank and Select~\cite{DBLP:conf/wads/BaumannH19}]\label{lem:rankselect}
Let $B$ be a bit string of length $m\le n$ stored in $\cO(1+m/\log n)$ words. We can preprocess $B$ in $\cO(1+m/\log n)$ time into a data structure of $m + o(m)$ bits  supporting \textsf{rank} and \textsf{select} queries in $\cO(1)$ time.    
\end{lemma}

Bit strings can also be used as a representation of monotonic integer sequences supporting predecessor queries; see~\cite{DBLP:journals/tcs/BadkobehCKP22}, for example. Assume we have a set $S$ of $m$ keys from a universe $U$ with a total order. In the \emph{predecessor} problem, we are given a query element $q\in U$, and we are to find the maximum $p\in S$ such that $p\leq q$; 
we denote this query by $\predecessor(q)=p$.
The following result is known for a special case of the predecessor problem.

\begin{lemma}[Fusion Tree~\cite{FW1993}]\label{lem:fusionnode} 
We can preprocess a set of $m=\log^{\cO(1)} n$ integers in $\cO(m)$ time and space to support $\predecessor$ queries in $\cO(1)$ time.
\end{lemma} 

\section{Constant-time Queries using \texorpdfstring{$\cO(n \log n)$}{} Space}\label{sec:constant_time}
We first show how to solve the \SWAQp problem in $\cO(1)$ time using $\cO(n \log n)$  space. This solution forms the basis for our linear-time and linear-space solution in \cref{sec:linear_space}. 

\subsection{Heavy-path Decomposition}
Let $T$ be a rooted tree with $n$ nodes.
We compute the \emph{heavy-path decomposition} of $T$ in $\cO(n)$ time~\cite{DBLP:journals/jcss/SleatorT83}. Recall that, for any node $u$ in $T$, we define
$\size(u)$ to be number of nodes in the subtree of $T$ rooted at $u$. We call an edge $(u,v)$ of $T$ \emph{heavy} if $\size(v)$ is maximal among every edge originating from $u$ (breaking ties arbitrarily). All other edges are called \emph{light}. We call a node that is reached from its parent through a heavy edge \emph{heavy}; otherwise, the node is called \emph{light}. The heavy path of $T$ is the path that starts at the root of $T$ and at each node on the path descends to the heavy child as defined above. The \emph{heavy-path decomposition} of $T$ is then defined recursively: it is a union of the heavy path of $T$ and the heavy-path decompositions of the off-path subtrees of the heavy path. A well-known property of this decomposition is that every root-to-node path in $T$ passes through at most $\log n$ light edges. In particular, the following lemma is implied.

\begin{lemma}[Heavy-path Decomposition~\cite{DBLP:journals/jcss/SleatorT83}]\label{lem:heavypath}
Let $T$ be a rooted tree with $n$ nodes. Any root-to-leaf path in $T$ consists of at most $\log n + \cO(1)$ heavy paths. 
\end{lemma}

\subsection{Data Structure}\label{sec:simpledatastructure}
We construct a heavy-path decomposition of $T$. Consider a heavy path $H = v_1 \ldots v_{\ell}$.
We construct a bit string $B(H)$ that represents the differences between node weights using unary coding. Suppose that nodes $v_1 \ldots v_{\ell}$ of $H$ are listed in decreasing order of their depth and let $\delta(v_i)=\weight(v_i)-\weight(v_{i-1})$, for all $i>1$. We define $B(H)$ as follows:
\[B(H)=\enc(\weight(v_1))\cdot \enc(\delta(v_2))\ldots \cdot \ldots \enc(\delta(v_i))\cdot\ldots\cdot \enc(\delta(v_{\ell})),\] 
where $\enc(i)$ denotes the unary code of $i$; i.e., $\enc(i)$ consists of $i$ $1$'s followed by a single $0$.
The important property of our encoding is that
the total number of $0$-bits in $B(H)$ is $\ell$ and the total number of $1$-bits is $\weight(v_{\ell})$. 

\begin{figure}[t]
    \centering
    \includegraphics[width=6cm]{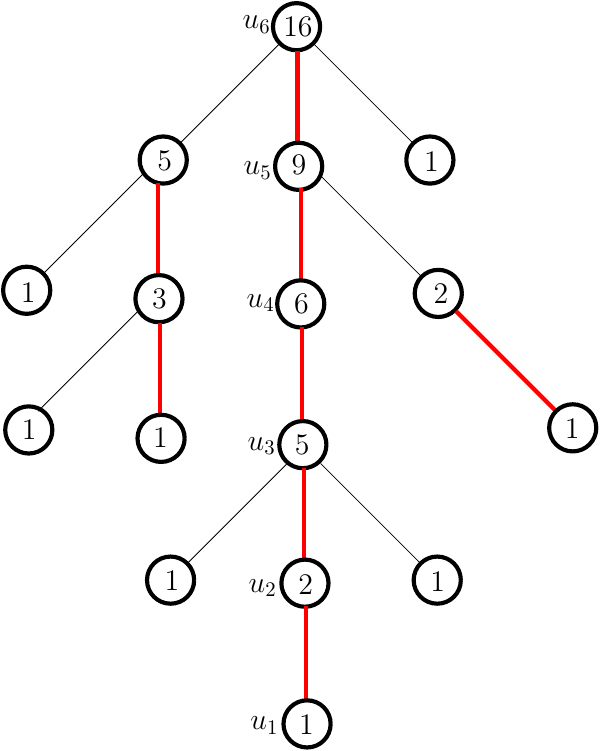}
    \caption{A rooted tree $T$ with $n=16$ nodes. Each node $u$ of $T$ is weighted by $\weight(u)= \size(u)$. 
    For example, $\weight(u_5)=\size(u_5)=9$, because there are $9$ nodes in the subtree rooted at $u_5$, and $\SWAQ(u_2,7)=u_5$ because the lowest ancestor of $u_2$ with weight at least $7$ is node $u_5$. A heavy-path decomposition of $T$ is also depicted: the heavy edges are the red edges. For example, the heavy path of the whole $T$ is $u_1u_2\ldots u_6$.}
    \label{fig:tree}
\end{figure}

\begin{example}\label{ex:running}
    Let $H=u_1u_2\ldots u_6$ be the heavy path of $T$ from Figure~\ref{fig:tree}. We have $\ell=6$ and
    $\weight(u_1)=1,~\weight(u_2)=2,~\weight(u_3)=5,~\weight(u_4)=6,~\weight(u_5)=9,~\weight(u_6)=16$.
    We have $B(H)=\texttt{1010111010111011111110}$.
    For instance, the second $1$ denotes $\delta(u_2)=\weight(u_2)-\weight(u_{1})=1$.
    The leftmost occurrence of $111$ denotes $\delta(u_3)=\weight(u_3)-\weight(u_{2})=3$ 1's.
\end{example}

For any heavy path $H$, we can construct $B(H)$ in $\cO(\ell)$ time using standard word RAM bit manipulations to construct the unary codes and concatenate the underlying bit strings.  By Lemma~\ref{lem:heavypath}, every leaf node of $T$ has $\cO(\log n)$ ancestors $v_t$, such that $v_t$ is the topmost node of some heavy path $H$. Since any node in $T$ is counted in the weight of $\cO(\log n)$ topmost nodes, the total weight of all topmost nodes, summed over all heavy paths $H$, is $\cO(n\log n)$. Thus, the total length of all bit strings $B(H)$ is $\cO(n \log n)$ and we can construct them all in $\cO(n)$ time since the total length of the heavy paths is $\cO(n)$. We store each such bit string according to Lemma~\ref{lem:rankselect} to support $\cO(1)$-time $\rank$ and $\select$ queries using $\cO(n)$ preprocessing time and words of space. Furthermore, for each leaf node $v$ in $T$ we store the weights of the top nodes of each heavy path on the path from the root to $v$. By Lemma~\ref{lem:heavypath}, there are $\cO(\log n)$ such top nodes for each leaf. For every leaf node we  store the weights of its top node ancestors in  a \emph{fusion tree} data structure according to Lemma~\ref{lem:fusionnode}. The total space used by  all such fusion trees is $\cO(n \log n)$ words and the preprocessing time is $\cO(n \log n)$. Finally, we construct a \emph{lowest common ancestor} (LCA) data structure over $T$. Such a data structure answers LCA queries in $\cO(1)$ time after $\cO(n)$-time and $\cO(n)$-space preprocessing~\cite{DBLP:conf/latin/BenderF00}.

\subsection{Queries}
\begin{figure}[t]
    \centering
    \begin{subfigure}{0.45\columnwidth}
    \centering
    \includegraphics[width=3cm]{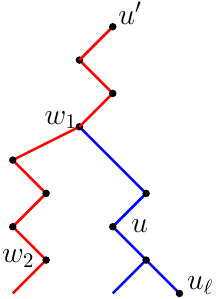}
    \caption{\textbf{Case 1}: Only $w_1$ is an ancestor of $u$. The heavy path $H_w$ is shown in red. The $(f+1)$th node $w_2$ on $H_w$ is below $w_1$. The node $w_1$ is the $(f+g)$th node on $H_w$ for some $g>1$, and so $w_1$ is the answer.} 
    \label{fig:query1}
    \end{subfigure}\hspace{+2mm}
    \begin{subfigure}{0.45\columnwidth}
    \centering
    \includegraphics[width=3cm]{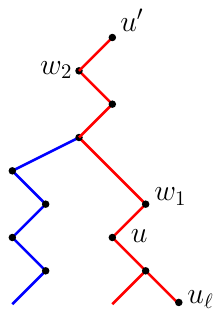}
    \caption{\textbf{Case 2}: Both $w_1$ and $w_2$ are ancestors of $u$. The heavy path $H_w$ is shown in red. The $(f+1)$th node $w_2$ on $H_w$ is above $w_1$, and so $w_2$ is the answer.\\} 
    \label{fig:query2}
    \end{subfigure}
    \caption{The two cases of the querying algorithm.}
    \label{fig:query}
\end{figure}

Suppose we are given a node $u$ and an integer $k$ as an $\SWAQ(u,k)$ query. We are looking for the lowest ancestor $w$ of $u$ with weight at least $k$. If the weight of $u$ is at least $k$, we return $u$. Otherwise we proceed as follows.  First, we locate the heavy path $H_w$ that contains node $w$: we find an arbitrary leaf descendant $u_\ell$ of $u$; then, using the fusion tree of $u_\ell$, we find the lowest ancestor $u'$ of $u_\ell$ with weight at least $k$, such that $u'$ is a top node. $H_w$ is the heavy path, such that $u'$ is its top node. When we find $H_w$,  we answer a query $f=\rank_0(B(H_w),j)$ for $j=\select_1(B(H_w),k)$ using Lemma~\ref{lem:rankselect} in $\cO(1)$ time.
Let $w_1$ denote the lowest ancestor of $u$ on the heavy path $H_w$ (see Figure~\ref{fig:query}). If $u$ is on $H_w$ (\cref{fig:query2}), then $w_1$ is simply the parent of $u$. Otherwise (\cref{fig:query1}), $w_1$ can be found as the lowest common ancestor of the lowest node on $H_w$ and node $u$. In the latter case, $w_1$ can be found using an LCA query that takes $\cO(1)$ time. Let $w_2$ denote the $(f+1)$th node on $H_w$. The node $w$ is the highest node among $w_1$ and $w_2$. The query time is $\cO(1)$ by Lemma~\ref{lem:fusionnode} for finding $H_w$ and by Lemma~\ref{lem:rankselect} for finding $f$. \cref{ex:rank_select} shows how we use $B(H_w)$ to find $f$ and thus the $(f+1)$th node. 

\begin{example}\label{ex:rank_select}
    Let $B(H)=\texttt{1010111010111011111110}$ from Example~\ref{ex:running}, $u_2$ from Figure~\ref{fig:tree}, and $k=7$.
    Then $j=\select_1(B(H),7)=11$ and
    $f=\rank_0(B(H),11)=4$. The output node is $u_5$, the $(f+1)$th node on $H$.
    Indeed, $\weight(u_5)=9\geq k = 7$ and $\weight(u_4)=6 < k = 7$.
\end{example}

In summary, we have shown the following result, which we will improve in the next section. 

\begin{lemma}\label{lem:swaqsimple}
For any rooted tree with $n$ nodes weighted by a size-constrained max-heap function $\weight$, there exists
an $\cO(n \log n)$-space  data structure answering
$\SWAQ$ queries in $\cO(1)$ time. The preprocessing algorithm runs in $\cO(n\log n)$ time and $\cO(n \log n)$ space.
\end{lemma}

\section{Constant-time Queries using \texorpdfstring{$\cO(n)$}{}  Space}\label{sec:linear_space}

We now improve the above solution to the \SWAQp problem (\cref{lem:swaqsimple}) to linear-time and linear-space preprocessing. We will reuse the previous section's linear-time heavy-path decomposition and the corresponding bit string encoding. The key challenge is identifying the top nodes of heavy paths in $\cO(1)$ time using linear space. 

\subsection{ART Decomposition}
The ART decomposition, proposed by Alstrup, Husfeldt, and Rauhe~\cite{AHR1998}, partitions a rooted tree into a \emph{top tree} and several \emph{bottom trees} with respect to an input parameter $\chi$. Each node $v$ of minimal depth, with no more than $\chi$ leaf nodes below it, is the root of a bottom tree consisting of $v$ and all its descendants. The top tree consists of all nodes that are not in any bottom tree. The ART decomposition satisfies the following important property: 

\begin{lemma}[ART Decomposition~\cite{AHR1998}]\label{lem:artdecomposition}
    Let $T$ be a rooted tree with $\ell$ leaf nodes. 
    Further let $\chi$ be a positive integer.
    The ART decomposition of $T$ with parameter $\chi$ produces a top tree with at most  $\cO(\ell/\chi)$ leaves.
    Such a decomposition of $T$ can be computed in linear time.
\end{lemma}

\subsection{Data Structure}
Recall that $T$ consists of $n$ nodes. As discussed in Section~\ref{sec:simpledatastructure}, we compute the heavy-path decomposition of $T$, construct bit strings for each heavy path, and preprocess the bit strings to support $\rank$ and $\select$ queries in $\cO(1)$ time. This takes $\cO(n)$ preprocessing time and space, allowing us to answer queries on a heavy path in $\cO(1)$ time. Thus what remains is a linear-space and $\cO(1)$-time solution to locate the top nodes of heavy paths. 

\begin{figure}[ht]
    \centering
    \begin{subfigure}{0.45\columnwidth}
    \centering
    \includegraphics[width=5cm]{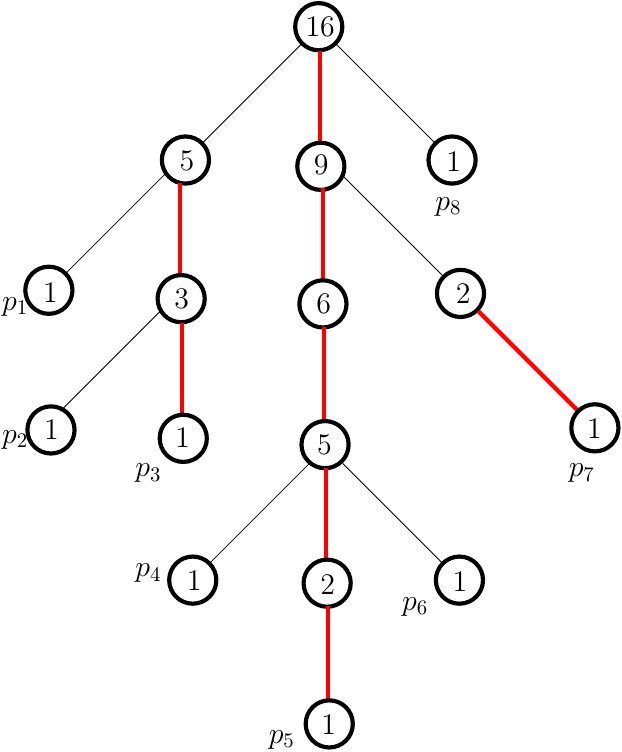}
    \caption{The tree $T$ from Figure~\ref{fig:tree}. We write the heavy path id $p_i$ at the end of the $i$th heavy path.} 
    \label{fig:path_tree}
    \end{subfigure}\hspace{+2mm}
    \begin{subfigure}{0.45\columnwidth}
    \centering
    \includegraphics[width=5cm]{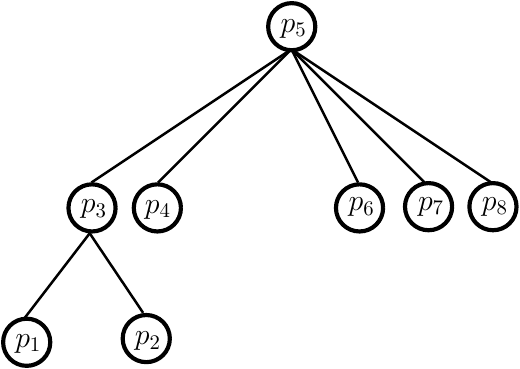}
    \caption{The contracted tree $C_T$.} 
    \label{fig:contracted_tree}
    \end{subfigure}
    \caption{The contraction process of the tree $T$ from Figure~\ref{fig:tree}.}
    \label{fig:contraction}
\end{figure}

\begin{figure}[ht]
    \centering
    \begin{subfigure}{0.45\columnwidth}
    \centering
    \includegraphics[width=6.5cm]{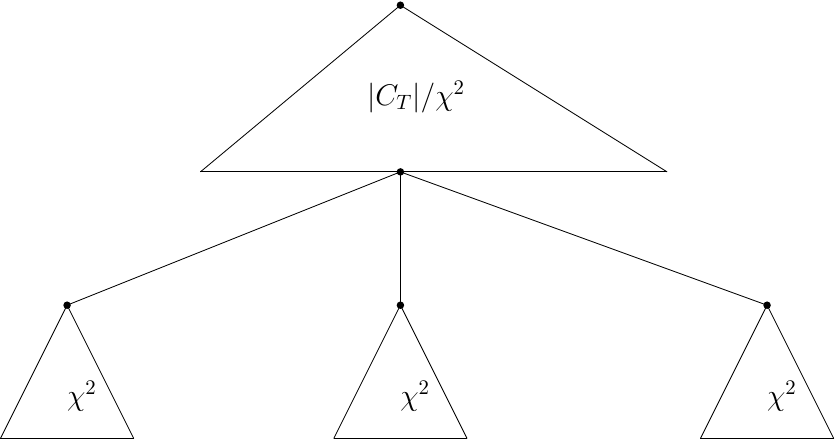}
    \caption{First application of ART decomposition on $C_T$.} 
    \label{fig:ART1}
    \end{subfigure}\hspace{+3mm}
    \begin{subfigure}{0.45\columnwidth}
    \centering
    \includegraphics[width=6.5cm]{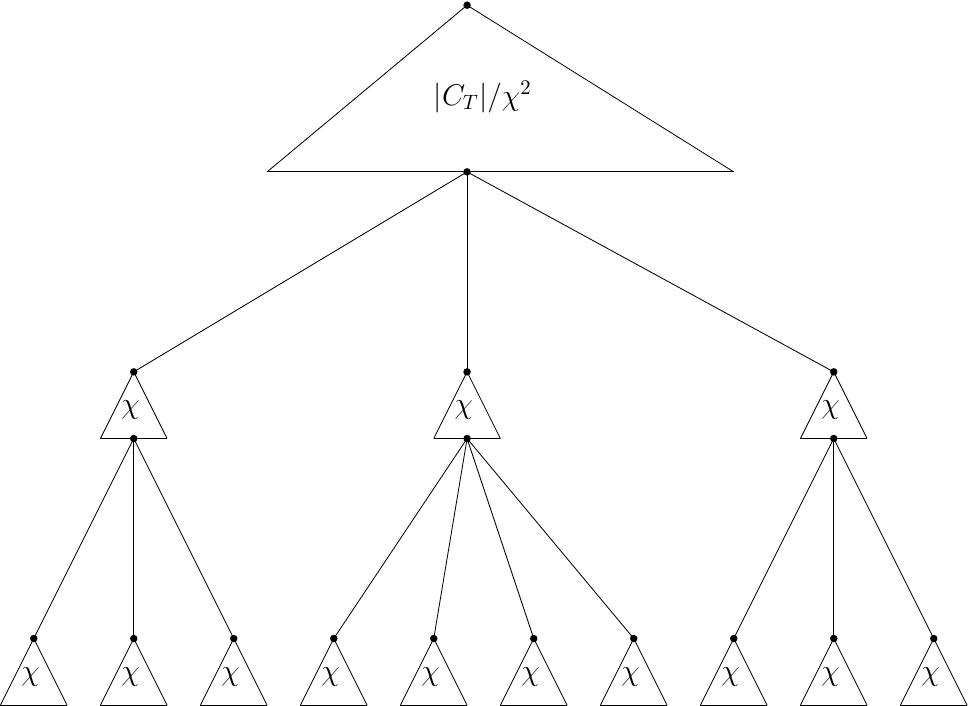}
    \caption{Application of ART decomposition on the bottom trees of the tree in Figure~\ref{fig:ART1}.} 
    \label{fig:ART2}
    \end{subfigure}
    \caption{Application of ART decompositions on $C_T$.}
    \label{fig:ART}
\end{figure}

First, we construct the \emph{contracted tree} $C_T$ of $T$ obtained by contracting all edges of heavy paths in $T$. 
In particular, this leaves all the light edges from $T$ in $C_T$ and removes all the heavy edges from $T$ (see Figure~\ref{fig:contraction}). We then apply the ART decomposition on $C_T$ (see Figure~\ref{fig:ART1}) with parameter $\chi^2$, where $\chi = \epsilon \frac{\log n}{\log \log n}$ and $\epsilon$ is a positive constant. We apply the ART decomposition again with parameter $\chi$ (see Figure~\ref{fig:ART2}) on each resulting bottom tree. The resulting partition of $C_T$ contains three levels of trees that we call the \emph{top tree}, the \emph{middle trees}, and the \emph{bottom trees}. 
Since the heavy-path decomposition of $T$ can be computed in $\cO(n)$ time, contracting $T$ takes $\cO(n)$ time by processing the heavy-path decomposition of $T$. By Lemma~\ref{lem:artdecomposition}, the ART decompositions of $T$ cost $\cO(n)$ total time. 

Let us first consider the top tree. As in Section~\ref{sec:simpledatastructure}, we store a fusion tree for each leaf node in the top tree. By Lemma~\ref{lem:artdecomposition}, the top tree has $\cO(\frac{|C_T|}{\chi^2})$ leaves and hence, by Lemmas~\ref{lem:fusionnode} and~\ref{lem:heavypath},  this uses $\cO(\frac{|C_T|}{\chi^2} \cdot \log n) = \cO(\frac{n (\log \log n)^2}{\log n}) = o(n)$ space and preprocessing time.   

For the middle or bottom trees, we tabulate the answers to all possible queries in a global table. The index in the table is given by a tree encoding and the node $u$ along with integer $k$ for the $\SWAQ$ query. The corresponding value in the table is the output node of the $\SWAQ(u,k)$ query. We encode the input to a query as follows. We represent each middle and bottom tree compactly as a bit string encoding the tree structure and the weights of all nodes. Since each internal node in $C_T$ is branching, the number of nodes in a middle or bottom tree is bounded by $\cO(\chi)$.
Thus, we can encode the tree structure using $\cO(\chi)$ bits. The weight of a node in a middle or bottom tree is bounded by $\cO(\chi^2)$ or $\cO(\chi)$, respectively, and can thus be encoded in $\cO(\log \chi)$ bits. 
Hence, we can encode the tree structure and all weights using $\cO(\chi \log \chi)$ bits. We encode the query node $u$ using $\cO(\log \chi)$ bits. Since the maximum weight is $\cO(\chi^2)$ we can also encode the query integer $k$ using $\cO(\log \chi)$ bits. Hence, the full encoding uses $\cO(\chi \log \chi) + \cO(\log \chi) + \cO(\log \chi) = \cO(\chi \log \chi)$ bits. To encode the output node stored in the global table we use $\cO(\log \chi)$ bits. Thus, the table uses $2^{\cO(\chi \log \chi)}\log \chi = 2^{\cO(\epsilon \log n)} = o(n)$ bits for a sufficiently small constant $\epsilon > 0$. The table can be constructed in $o(n)$ time.

\subsection{Queries}
Suppose we are given a node $u$ and an integer $k$ as an $\SWAQ(u,k)$ query. Let $u_t$ denote the top node on the heavy path of $u$ in $T$ and let $u_H$ denote the corresponding node in the contracted tree $C_T$.  We find the lowest ancestor $w_H$ of $u_H$ with weight at least $k$ in $C_T$. If $u_H$ is in the top tree we find $w_H$ as described in Section~\ref{sec:simpledatastructure}. If $u_H$ is in a middle or bottom tree, we use the global table to find $w_H$. If the result is not in the middle or bottom tree (the weight of the top node in such a tree is smaller than $k$), we move up a level and query the middle or top tree, respectively. Each of these at most three queries takes $\cO(1)$ time. Thus $w_H$ is found in $\cO(1)$ time. 
Suppose that $w_H$ corresponds to a node $w'$ in the initial tree and let $H'$ denote the heavy path such that $w'$ is its top node. As explained in Section~\ref{sec:simpledatastructure}, we can find the lowest ancestor of $u$ with weight at least $k$  on $H'$ in  $\cO(1)$ time using rank and select queries on $B(H')$. 
In total the $\SWAQ(u,k)$ query takes $\cO(1)$ time. 

In summary, we have obtained the following result. 

\theoremmain*

\section{String-processing Applications}\label{sec:apps}

In this section, we show several applications of Theorem~\ref{the:swaq} on suffix trees.
Recall that the number of leaf nodes in the subtree rooted at node $u$ in a suffix tree is the number of occurrences (i.e., the frequency) of the substring represented by the root-to-$u$ path.

\subsection{Internal Longest Frequent Prefix}\label{sec:ILFP}

Internal pattern matching is an active topic~\cite{DBLP:conf/soda/KociumakaRRW15,DBLP:journals/algorithmica/AmirCPR20,DBLP:journals/algorithmica/Charalampopoulos21,DBLP:conf/cpm/Charalampopoulos20,DBLP:conf/spire/CrochemoreIRRSW20,DBLP:journals/algorithms/Abedin0PT20,DBLP:journals/tcs/BadkobehCKP22} in the combinatorial pattern matching community.
We introduce the following basic string problem.
The \emph{internal longest frequent prefix} problem asks to preprocess a string $X$ of length $n$ over an integer alphabet $\Sigma=[1,n^{\cO(1)}]$ into a compact data structure supporting the following on-line queries: 
\begin{itemize}
    \item $\mathsf{ILFP}_X(i, j, f)$: return the longest prefix of $X[i\dd j]$ occurring at least $f$ times in $X$. 
\end{itemize}

Our solution to this problem will form the basic tool for solving the problems in \cref{sec:LFS,sec:FC}.
We first construct the suffix tree $T$ of $X$ in $\cO(n)$ time~\cite{DBLP:conf/focs/Farach97},
and preprocess it in $\cO(n)$ time for classic weighted ancestor queries~\cite{DBLP:conf/cpm/BelazzouguiKPR21} as well as for $\SWAQ$ queries using Theorem~\ref{the:swaq}.
For $\SWAQ$ queries, as $\weight(u)$, we use the number of leaf nodes in the subtree rooted at node $u$ in $T$. Such an assignment satisfies the requested properties of $\weight(\cdot)$ and can be done in linear time using a standard DFS traversal on $T$. Any $\mathsf{ILFP}_X(i, j, f)$ query can be answered by first finding the locus $(u,j-i+1)$ of $X[i\dd j]$ in $T$ in $\cO(1)$ time using a classic weighted ancestor query on $T$, and, then, answering $\SWAQ(u,f)$ in $T$ in $\cO(1)$ time using Theorem~\ref{the:swaq}. We obtain the following result.

\begin{theorem}
  For any string $X$ of length $n$ over alphabet $\Sigma=[1,n^{\cO(1)}]$, there exists an $\cO(n)$-space data structure that answers $\mathsf{ILFP}_{X}$ queries in $\cO(1)$ time. The preprocessing algorithm runs in $\cO(n)$ time and $\cO(n)$ space.   
\end{theorem}

\subsection{Longest Frequent Substring}\label{sec:LFS}

The \emph{longest frequent substring} problem is the following: preprocess a dictionary $\mathcal{D}$ of $d$ strings (documents) of total length $n$ over an integer alphabet $\Sigma=[1,n^{\cO(1)}]$ into a compact data structure supporting the following on-line queries:

\begin{itemize}
    \item $\mathsf{LFS}_{\mathcal{D}}(P, f)$: return a longest substring of $P$ that occurs in at least $f$ documents of $\mathcal{D}$. 
\end{itemize}

This longest substring of $P$ represents a most \emph{relevant} part of the query with respect to  $\mathcal{D}$. The length of $\mathsf{LFS}_{\mathcal{D}}(P, f)$ can also be used as a \emph{measure of similarity} 
between $P$ and the strings in $\mathcal{D}$, for some $f$ chosen appropriately based on the underlying application.

We start by constructing the generalized suffix tree $T$ of $\mathcal{D}$ in $\cO(n)$ time~\cite{DBLP:conf/focs/Farach97}
and preprocess it in $\cO(n)$ time for $\SWAQ$ queries using Theorem~\ref{the:swaq}.
For $\SWAQ$ queries, $\weight(u)$ is equal to  the number of
dictionary strings having at least one leaf node in the subtree rooted at node $u$ in $T$.
This assignment satisfies the requested properties of $\weight(\cdot)$ and can be done in linear time~\cite{DBLP:conf/cpm/Hui92}.
Let us denote by $(v_i,\ell_i)$ the locus in $T$ of the longest prefix of $P[i\dd |P|]$ that occurs in any string in $\mathcal{D}$. In fact, we can compute $(v_i,\ell_i)$, for all $i\in [1,|P|]$, in $\cO(|P|)$ time using the \emph{matching statistics} algorithm of $P$ over $T$~\cite{DBLP:journals/algorithmica/ChangL94,DBLP:books/cu/Gusfield1997}. For each locus $(v_i,\ell_i)$, we trigger a $\SWAQ(v_i,f)$ query using Theorem~\ref{the:swaq} (this is essentially an instance of the \emph{internal longest frequent prefix} problem). 
In total this takes $\cO(|P|)$ time.
We obtain the following result.

\begin{theorem}\label{the:LFS}
  For any dictionary $\mathcal{D}$ of total length $n$ over alphabet $\Sigma=[1,n^{\cO(1)}]$, there exists an $\cO(n)$-space data structure that answers $\mathsf{LFS}_{\mathcal{D}}(P, f)$ queries in $\cO(|P|)$ time.
  The preprocessing algorithm runs in $\cO(n)$ time and $\cO(n)$ space.
\end{theorem}

An analogous result can be achieved for the following version of the longest frequent substring problem: preprocess a string $X$ of length $n$ over an integer alphabet $\Sigma=[1,n^{\cO(1)}]$ into a compact data structure supporting the following on-line queries:

\begin{itemize}
    \item $\mathsf{LFS}_{X}(P, f)$: return a longest substring of $P$ that occurs at least $f$ times in $X$. 
\end{itemize}

In particular, instead of a generalized suffix tree, we now construct the suffix tree $T$ of $X$ and follow the same querying algorithm as above. For $\SWAQ$ queries, $\weight(u)$ is equal to  the number of leaf nodes in the subtree rooted at node $u$ in $T$. Such an assignment satisfies the requested properties of $\weight(\cdot)$ and can be done in linear time using a standard DFS traversal on $T$. We obtain the following result.

\begin{theorem}\label{the:LFS_X}
    For any string $X$ of length $n$ over alphabet $\Sigma=[1,n^{\cO(1)}]$, there exists an $\cO(n)$-space data structure that answers $\mathsf{LFS}_{X}(P, f)$ queries in $\cO(|P|)$ time. The preprocessing algorithm runs in $\cO(n)$ time and $\cO(n)$ space.
\end{theorem}

\subsection{Frequency-constrained Substring Complexity}\label{sec:FC}

For a string $X$, a dictionary $\mathcal{D}$ of $d$ strings (documents) and a partition of $[d]$ in $\tau$ intervals $\mathcal{I}=I_1,\ldots,I_\tau$, the function $f_{X,\mathcal{D},\mathcal{I}}(i,j)$ maps $i,j$ to the number of distinct substrings of length $i$ of $X$ occurring in at least $\alpha_j$ and at most $\beta_j$ documents in $\mathcal{D}$, where $I_j=[\alpha_j,\beta_j]$. 
Function $f$ is known as the
\emph{frequency-constrained substring complexity} of $X$~\cite{DBLP:conf/spire/PissisSLL23}.

\begin{example}
Let $\mathcal{D}=\{\texttt{a,ananan,baba,ban,banna,nana}\}$. For $X=\texttt{banana}$ and $I_1=[1,2],I_2=[3,4],I_3=[5,6]$, we have $f_{X,\mathcal{D},\mathcal{I}}(2,2)=3$: $\texttt{ba}$ occurs in $3 \in I_2$ documents; $\texttt{an}$ occurs in $4 \in I_2$ documents; and $\texttt{na}$ occurs in $3 \in I_2$ documents. 
\end{example}

The function $f_{X,\mathcal{D},\mathcal{I}}$ is very informative about $X$; it provides fine-grained information about the contents (the substrings) of $X$. It can thus facilitate the tuning of string-processing algorithms by setting bounds on the length or on frequency of substrings; see~\cite{DBLP:conf/spire/PissisSLL23}. 

Let $S$ be a 2D array such that $S[i,j]=f_{X,\mathcal{D},\mathcal{I}}(i,j)$. Pissis et al.~\cite{DBLP:conf/spire/PissisSLL23} showed that after an $\cO(n)$-time preprocessing of a dictionary $\mathcal{D}$ of $d$ strings of total length $n$ over an integer alphabet $\Sigma=[1,n^{\cO(1)}]$, for any $X$ and any partition $\mathcal{I}$ of $[d]$ in $\tau$ intervals given on-line, $S$ can be computed in near-optimal $\cO(|X| \tau\log\log d)$ time. 

The solution in~\cite{DBLP:conf/spire/PissisSLL23} can be summarized as follows. In the preprocessing step, we construct the generalized suffix tree $T$ of $\mathcal{D}$.
In querying, the first step is to construct the suffix tree of $X$ and compute the document frequency of its nodes in $\cO(|X|)$ time. In the second step, we enhance the suffix tree of $X$ with $\cO(|X|\tau)$ nodes with document frequencies by answering $\SWAQ$ queries on $T$ in $\cO(\log\log d)$ time per query~\cite{DBLP:journals/talg/AmirLLS07}. The whole step thus takes $\cO(|X| \tau\log\log d)$ time. In the third step, we infer a collection of length intervals, one per node of the enhanced suffix tree and sort them in $\cO(|X|\tau)$ time using radix sort. In the last step, we sweep through the intervals from left to right to compute array $S$ in $\cO(|X|\tau)$ total time. 
This concludes the summary of the solution in~\cite{DBLP:conf/spire/PissisSLL23}.
We amend the solution as follows. 

We plug in Theorem~\ref{the:swaq} for preprocessing $T$ and for the second step ($\SWAQ$ queries). For $\SWAQ$ queries, as $\weight(u)$, we use the number of
dictionary strings having at least one leaf node in the subtree rooted at node $u$ in $T$.
Such an assignment satisfies the requested properties of $\weight(\cdot)$ and can be done in linear time~\cite{DBLP:conf/cpm/Hui92}. We obtain the following result.

\begin{theorem}\label{the:tableS}
For any dictionary $\mathcal{D}$ of $d$ strings of total length $n$ over alphabet $\Sigma=[1,n^{\cO(1)}]$, there exists an $\cO(n)$-space data structure that answers $S=f_{X,\mathcal{D},\mathcal{I}}$ queries in $\cO(|X|\tau)$ time. The preprocessing algorithm runs in $\cO(n)$ time and $\cO(n)$ space.
\end{theorem}

Since $S$ is of size $|X|\cdot\tau$ (it consists of $|X|\cdot\tau$ integers), the complexity bounds are optimal with respect to the preprocessing and query times.

\bibliography{references}

\end{document}